\documentclass{rspublic}

\usepackage{graphics}

\begin{document}
\title{Brane Inflation and Defect Formation}
\author[A.-C. Davis, Ph. Brax and C. van de Bruck] {Anne-Christine Davis$^1$, Philippe Brax$^2$ and Carsten van de Bruck$^3$}
\affiliation{$^1$DAMTP, Centre for Mathematical Sciences, \\
University of Cambridge, Cambridge CB3 0WA, UK \\
$^2$CEA, DSM, Institut de Physique Théorique, IPhT, CNRS, MPPU,
URA2306,
Saclay, F-91191 Gif-sur-Yvette, France\\
$^3$Department of Applied Mathematics, School of Mathematics and Statistics \\
Hounsfield Road, Sheffield S3 7RH, UK}

\label{firstpage}

\maketitle

\begin{abstract}{Brane inflation and cosmic superstrings}
Brane inflation and the production of topological defects at the
end of the inflationary phase are discussed. After a description
of the inflationary setup we discuss the properties of the cosmic
strings produced at the end of inflation. Specific examples of
brane inflation are described: $D-\bar D$ inflation, $D3/D7$
inflation and modular inflation.
\end{abstract}

\section{Introduction}

Superstring theory is so far our best hope for a theory which
unifies quantum mechanics, gravity and the other known forces in
nature (see e.g. Becker et al. (2006) and Polchinski (1998) for an
overview of string theory). Until recently it has been very
difficult to conceive tests of stringy predictions, i.e. departure
from usual particle physics, which are not all at very high energy
scales close to the Planck scale. However, it was recently realized
that cosmological considerations in string theory lead to verifiable
predictions, testable with current and future cosmological
experiments such as WMAP and PLANCK. The reason for this is twofold.
On the one hand, theoretical developments have led to string
predictions at energy scales less than the Planck scale due to a
method of compactification with fluxes. On the other hand
cosmologists now believe that the observed structures in the
universe such as galaxies and clusters of galaxies have their origin
in the very early universe (see e.g. Mukhanov (2006) for an
overview). Most cosmologists believe that the seeds of these
structures were created during a period of rapid exponential
expansion, called inflation.

Inflation was proposed to solve a number of cosmological problems in
the early Universe, such as the flatness problem, the horizon
problem and the overproduction of defects. It predicts an almost
scale-invariant power spectrum and temperature anisotropies
in excellent agreement with the recent maps from satellite
experiments of COBE and WMAP (see e.g. Liddle \& Lyth (2000) and
Mukhanov (2006)). A fundamental issue is to find realistic models of
inflation within particle physics theories.

On the other hand, if string theory really is the theory of
everything, then it would be natural to search for a model of
inflation within string theory. A number of recent developments
within string theory have enabled one to start to build a model of
inflation, called brane inflation (see e.g. Dvali \& Tye (1998),
Kachru et al. (2003b), Burgess et al. (2004) and references
therein). In the following we will describe the idea of brane
inflation and discuss the formation of topological defects in these
models.

The outline  of this article is as follows. We first give a
general introduction to brane inflation in string theory. Brane
inflation results in the formation of lower dimensional branes,
and in particular D-strings. We discuss the properties of
D-strings in the low energy supergravity description of brane
inflation. Finally we discuss the possibility that cosmic string
formation at the end of brane inflation may not be generic in the
context of modular inflation.

\section{Brane Inflation}

The number of string inflation models is rather large, hence we
will focus on a limited number of examples. All these examples
will have a supergravity counterpart and are describable by a
field theory at low energy. There are two types of relevant
models. The first one is the ${\rm D3}-\overline{\rm D3}$ system.
The second one is the $D3/D7$ system; both realise  hybrid
inflation in string theory.

Before delving  into the presentation of these systems, let us
recall some basic notions about branes in string theory (see Becker et al
(2006)). String theory is formulated in $10$ space-time
dimensions, $9$ space and $1$ time. It deals with closed and open
strings. Open string are strands ending on extended objects called
branes, i.e. their extremities are attached to submanifolds of the
ten dimensional space-time. Type IIB string theory admits branes of odd
spatial dimensions. After reducing space-time on a six dimensional
manifold in order to retrieve our four dimensions of space-time,
only D3 and D7 branes give supersymmetric (which is a symmetry between
fermions and bosons) configurations. Anti D-branes preserve opposite
supersymmetries and therefore break all supersymmetries in the presence of
branes. A ${\rm D3}-\overline{\rm D3}$ system breaks all supersymmetries
explicitly while a $D3/D7$ system preserves supersymmetry. The
compactification process on a six dimensional manifold introduces
a host of massless fields called moduli. They parameterise all the
possible deformations of the six-manifold. The existence of massless moduli
would have a catastrophic effect in the low energy description of string
theory: they would imply a strong modification of Newton's law of
gravity. To avoid this, moduli must become
massive and therefore acquire a potential which stabilises their
value. The potential for the moduli has been obtained recently in
a scenario called the KKLT approach, see Kachru et al (2003a). It involves two
new ingredients. The first one are fluxes which resemble constant
magnetic fields created by wires. These fluxes stabilise half of
the moduli leaving only the moduli measuring the size of the
six-manifold untouched. They also imply that the compactification
manifold is warped, i.e. looks like an elongated throat attached
to a region called the bulk, as shown in fig 1. When space-time is warped
the invariant distance becomes
\begin{equation}
ds^2 = e^{-A(y)}(dt^2 - dx^2) -dy^2
\end{equation}
where $A(y)$ is the warp factor and $y$ represents the extra dimensions.
At the bottom of the throat, energy
levels are red-shifted and can be much lower than the string
scale. The size moduli can be stabilised thanks to non-perturbative effects
on D7 branes. Unfortunately, the stabilisation is realised
with a negative vacuum energy, ie anti-de Sitter space-time.
It is necessary to lift this up to Minkowski space-time with
an `uplifting' mechanism.

\begin{figure}
\begin{center}
\scalebox{0.7}{\includegraphics{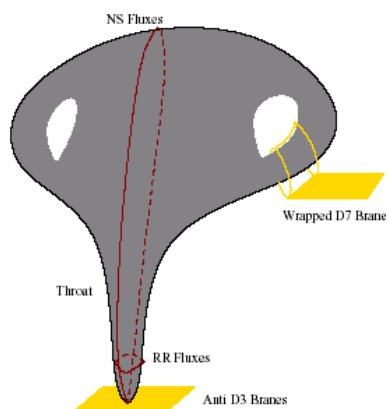}}
\end{center}
\caption{A throat generated by the fluxes on the compactification
manifold. An anti-brane sitting at the bottom of the throat is shown
while another D-brane, where non perturbative effects can take place,
has also been depicted. }
\end{figure}

Let us first discuss inflation in the ${\rm D3}-\overline{\rm D3}$
system (see Kachru et al (2003b)). First of all, the $\overline{\rm D3}$ has a
very specific role in this model. It gives the positive energy
which is necessary to lift up the negative minimum of the moduli
potential. Inflation is obtained after introducing a D3 brane
which is attracted towards the $\overline{\rm D3}$ brane.  The
inflaton field is the distance between the branes. As the branes
get closer, an instability develops in a similar way to hybrid
inflation, with potential as in fig 2. This model is a nice description of
hybrid inflation in string theory and one of the main achievements of string
inflation: providing a fundamental description of the inflaton
field. There are however problems with this. As a paradigm, the ${\rm
D3}-\overline{\rm D3}$ system has a few unambiguous consequences:
the level of primordial gravitational waves is very small and
cosmic strings with a very low tension may be produced

\begin{equation}
\mu = e^{-A(y)} \mu_0,
\end{equation}
where $\mu_0$ is the fundamental string tension.
Nevertheless, the lack of a proper
supersymmetric setting for the $D3-\overline{D3}$ renders it less
appealing than the D3/D7 system that we are about to discuss.

The D3/D7 system has two big advantages over the ${\rm
D3}-\overline{\rm D3}$ systems: supersymmetry and it contains a shift symmetry
which eliminates the so--called $\eta$ problem, see Dasgupta et al. (2004). In the D3/D7
system, the inflaton is the inter-brane distance while there are
waterfall fields corresponding to open string joining the branes.
It is a clean stringy realisation of D-term inflation. In this
context, the existence of Fayet-Iliopoulos terms is crucial. They
are needed to lift the moduli potential and to provide the
non-trivial value of the waterfall fields at the end of inflation.

\begin{figure}
\begin{center}
\scalebox{0.7}{\includegraphics{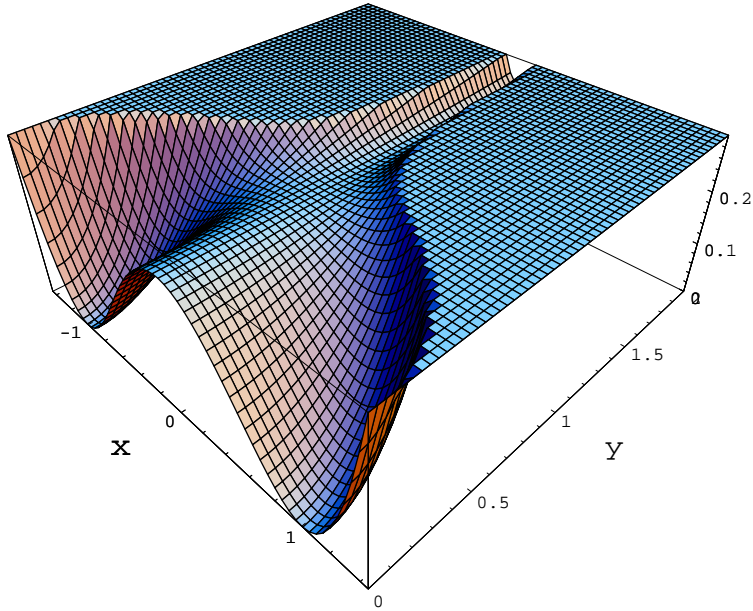}}
\end{center}

\caption{The potential
for hybrid inflation. The inflaton field rolls down the almost flat
direction until an instability results. The waterfall fields then
acquire their vacuum values, resulting in the end of inflation.}
\end{figure}

They can be obtained using magnetic fields on $\rm D7$ branes. The
condensation of the waterfall fields is the equivalent of the
stringy phenomenon whereby a $D3$ brane dissolves itself into a $D7$
brane, leaving as remnants an embedded $D1$ brane. These branes are
one dimensional objects differing from fundamental strings. It has
been conjectured and there is evidence in favour of the fact that
these D-strings correspond to the BPS D-term strings of the low
energy supergravity description. Unfortunately, it turns out that
the tension of these strings is quite high implying that they
would lead to strong features in the CMB spectra. At the moment,
the problem of both generating a low tension D-string and
inflation at the relevant scale, given by the COBE normalisation,
has not been solved. Including more $D7$ branes gives semi-local strings,
see Urrestilla et al (2004), which alleviate the problem.
On the positive side, these models have the
nice feature of giving $n_s\approx 0.98$ which could be compatible
with data. On the other hand, the spectrum of gravitational waves
is undetectable.

The ${\rm D3}-{\overline {\rm D3}}$ and $D3/D7$ systems are
promising avenues to obtain realistic string inflation models. At
the moment, they are no more than plausible scenarios. Inflation
being a very high scale phenomenon, it is clear that the study of
early universe phenomena could be a crucial testing ground for
string theory.

\section{Cosmic D-strings}
The formation of cosmic strings appears to be a generic feature of
recent models of brane inflation arising from fundamental string
theory, see e.g. Sarangi \& Tye (2002) and Copeland et al (2004),
for a review see Davis \& Kibble (2005), Polchinski (2004). Indeed,
lower dimensional branes are formed when a brane and anti-brane
annihilate with the production of $D3$ and $D1$ branes, or
D-strings, favoured (see Majumdar \& Davis (2002)). It has been
argued that D-strings have much in common with cosmic strings in
supergravity theories and that they could be identified with D-term
strings, see Dvali et al (2004). This is because D-strings should be
minimum energy solutions known as BPS solutions. The D-term strings
in supergravity are such BPS solutions. Not only are BPS strings
minimum energy, but the BPS property means that the equations of
motion, which should be second order differential equations, reduce
to first order equations making them simpler to solve.

Whilst cosmic strings in global
supersymmetric theories have been analysed
before, see Davis et al(1996) the study of such strings in supergravity
is at an early stage since there are added complications.
In Brax et al (2006) an exhaustive study of fermion zero modes was performed.
In supersymmetry there are fermions as well as bosons in the action. The
fermion partners of the bose string fields can be zero modes. These are
zero energy solutions of the Dirac equation. It was found
that due to the presence of the gravitino, the number of zero mode
solutions in supergravity is reduced in some cases.
For BPS D-strings with winding number $n$ the number of chiral zero
modes is  $2(n-1)$, rather than $2n$ in the global case. When there are
spectator fermions present, as might be expected if the underlying theory
gives rise to the standard model of particle physics, then there are $n$
chiral zero modes per spectator fermion.

Following Dvali et al (2004) we use the supergravity description of
D-strings by D-term strings. Consider a supergravity
theory with fields $\phi^\pm$, charged under an Abelian gauge
group. The D-term bosonic potential
\begin{equation}
V = \frac{g^2}{2}(|\phi_+|^2 - |\phi_-|^2 - \xi)^2
\end{equation}
includes a non-trivial Fayet-Iliopoulos term $\xi$. Such a term is
compatible with supergravity provided the superpotential has
charge $-\xi$. Here the superpotential vanishes.  The minimum of
the potential is $|\phi_+|^2=\xi$. It is consistent to take the
cosmic string solution to be
\begin{equation} \phi_+ =
f(r) e^{in\theta}
\end{equation}
where $n$ is the winding number, and $\phi_-= 0$. The function $f(r)$
interpolates between $0$ and $\sqrt \xi$. The presence of a cosmic
string bends space-time. Its gravitational effects
lead to a deficit angle in the far away metric of
spacetime. In the following we consider the metric
\begin{equation}
ds^2=e^{2 B}(-dt^2 + dz^2) + dr^2 + C^2 d\theta^2
\end{equation}
for a cosmic string configuration. This is the most general
cylindrically symmetric metric. Far away from the string the energy momentum
tensor is approximately zero and
therefore $B$ is zero. Now we also find that
\begin{equation}
C =  C_1 r + C_0 + O(r^{-1}) \ ,
\end{equation}
When $C_1\ne 0$, the solution is a cosmic string solution with a
deficit angle $\Delta= 2\pi (1-C_1)$.

In supersymmetric theories, a cosmic string generally breaks all
supersymmetries in its core. BPS objects are an exception
to this rule, as they leave 1/2 of the original supersymmetry
unbroken. D-strings, in which we will be interested in this paper,
are an example of this. These strings have vanishing $T_{rr}$,
and the conformal factor $B$ is identically zero. Moreover for
D-strings one finds
\begin{equation}
\Delta = 2 \pi |n|\xi \ .
\end{equation}
 The supersymmetry algebra in four dimensions allows for 1/2
BPS configurations which saturate a BPS bound giving an equality
between the mass, i.e.\ the tension, and a central charge. Other
cosmic strings have higher tension and are not BPS, i.e.\ they
break all supersymmetries. This implies that $C_1$ is less than
the BPS case giving
\begin{equation}
\Delta\ge 2\pi\vert n \vert \xi \ .
\end{equation}

Let us characterise the BPS cosmic strings. We are considering a
$U(1)$ symmetry breaking, so we take its generator $T_s \phi^i = n
Q_i \phi^i$ and $A_\mu = \delta^\theta_\mu n a(r)$. The bosonic
fields satisfy first order equations
\begin{equation}
\partial_r \phi^i = \mp n\frac{1-a}{C} Q_i \phi^i
\label{Dstr1}
\end{equation}
and
\begin{equation}
\mp n \frac{\partial_r a}{C} = D = \xi - \sum_i Q_i K_i \bar \phi^i
\end{equation}
where $\xi$ is the Fayet-Iliopoulos term which triggers the breaking of
the $U(1)$ gauge symmetry and $K$ is the Kahler potential. The
Einstein equations reduce to $B'=0$ and
\begin{equation}
C'=1\pm A^B_\theta
\label{Dstr3}
\end{equation}
where
\begin{equation}
A^B_\mu = \frac{i}{2} (\bar K_{\bar \jmath} D_\mu \bar \phi^{\bar
\jmath} - K_j D_\mu \phi^j) +\xi A_\mu
\label{AB}
\end{equation}
The simplest BPS configuration will just have one $\phi$, with unit
charge and $K = \phi^+\phi^-$. Notice that BPS cosmic strings are solutions
of first order differential equations. These equations are consequences of the
Killing spinor equations when requiring the existence of 1/2
supersymmetry.

\section{A Specific Realisation of Brane Inflation}

As mentioned previously, a specific realisation of brane inflation is
obtained  in the system with a $D3$ and $D7$ brane, giving rise to
the formation of D-strings at the end of inflation. Here we use
the KKLT model with superpotential
\begin{equation}
W = W_0 + A e^{-aT}
\end{equation}
where $W_0$ arises after integrating out the other moduli and the
exponential term arises from non-perturbative effects.
In order to keep supersymmetry we can uplift with a so-called
Fayet-Iliopoulos term. This is more
complicated in supergravity than in ordinary supersymmetry. However,
it can be achieved using a method to ensure any anomalies are
cancelled, the so-called Green-Schwarz mechanism, see Achucarro et al (2006).
This gives us a more complicated superpotential
\begin{equation}
W_{\rm mod}(T,\chi) = W_0 + \frac{A e^{-a T}}{\chi^b} \, .
\label{Wmod}
\end{equation}
where $\chi$ is a field living on $D7$ branes. Following Brax et al (2007a), we
use a no-scale Kahler potential and include the inflationary terms arising
from the superstrings between the branes,
\begin{equation}
W_{\rm inf} = \lambda \phi \phi^+ \phi^-.
\label{Winf}
\end{equation}
we get potential $V = V_F + V_D$, where
\begin{equation}
V_D = \frac{g^2}{2}\left(|\phi^+|^2 - |\phi^-|^2 -
\xi \right)^2
\end{equation}
for the D-term and the F-term is calculated from the superpotential. The
Fayet-Iliopoulos term, $\xi$ is obtained from anomaly cancellation and
depends on the fields, $\chi$ and $T$. The scalar potential is quite
complicated, consisting of the part from the superpotential, which is called
the F-term part, and the D-term part. We have plotted them in fig 3, where
we can see that the original F-term part has a minimum that is less than
zero, but the combined potential has a minimum at zero when one adds in
the contribution of $V_D$.

\begin{figure}
\scalebox{0.5}{\includegraphics{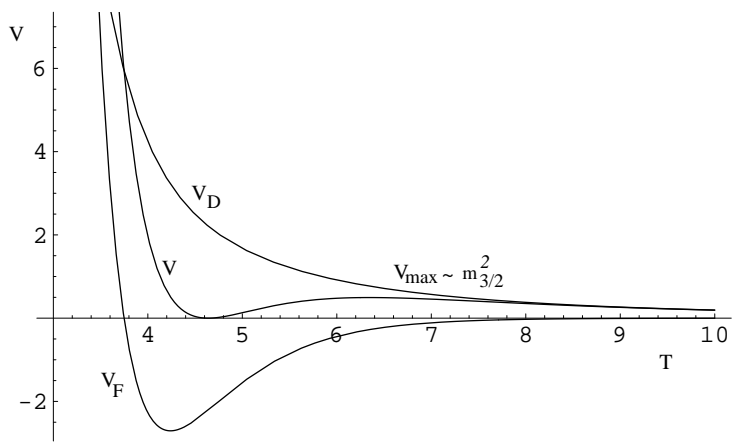}}
\caption{The potential for our model. Note that the F-term gives an AdS minimum, which is uplifted to Minkowski space with the D-term}
\end{figure}

Inflation proceeds as in fig 2 with the charged $\phi^\pm$ fields
being zero, but $\phi$ field being non-zero until the D$3$ brane
gets close to the D$7$. The charged fields become tachyonic,
acquiring a vacuum expectation value. They are called `waterfall'
fields since the fields fall to their minimum values. At this
point cosmic superstrings, or D-strings, are formed. We have
analysed the inflation in this theory and found that it gives the
density perturbations and spectral index in agreement with the
data from WMAP3. We have also analysed the string solutions formed
at the end of inflation. In our model of $D$-term inflation
coupled to moduli, cosmic strings form during the breaking of the
Abelian gauge symmetry $U(1)$ by the waterfall field $\phi^+$ at
the end of inflation.  The cosmic strings in our model are not BPS
as in usual D-term inflation.  This is due to the non-vanishing of
the gravitino mass and the coupling of the string fields to the
moduli sector. We find the  cosmic string solution has the form
\begin{equation}
\phi^+ = {\xi}^{1/2} e^{in\theta} f(r) \, , \qquad \phi^{-} = \phi
=0 \, , \qquad A_{\theta} = n a(r)
\end{equation}
and the moduli sector fields depend only on $r$. The functions
$f(r)$, $a(r)$, differ from  the standard ones appearing in the
Abelian string model (see e.g. Vilenkin \& Shellard (2000)) as
include the effect of contributions from the moduli field and
$\chi$ arising from the full potential. The Higgs field tends to
its vacuum value at infinity, and regularity implies it is zero at
the string core. Without loss of generality we can take the
winding number positive $n>0$.  If the variation of the moduli
fields inside the string is small, the string solution will be
similar to a BPS $D$-string, with a similar string tension, which
(for $n=1$) is
\begin{equation}
\mu \approx 2\pi \xi \, .
\end{equation}
The above result will receive corrections from the one loop
contribution to the potential. We also expect that the variation of
the moduli fields inside the string will reduce $\mu$.

\section{Is Cosmic String Formation Generic?}
There have been recent attempts to use the string moduli fields
themselves as the inflaton field. This type of model is known as
racetrack inflation, see Blanco-Pillado et al (2004) and Brax et
al (2007b). One can either use an $\bar{D}$ brane or a D-term for
uplifting, as in the previous section. In this class of theory
there are no matter fields since the open string modes between
branes are not included. Consequently there are no cosmic strings
produced at the end of inflation. However, this is a rather
artificial situation. Here we will describe a preliminary attempt
to include the matter fields.

The superpotential for racetrack inflation is
\begin{equation}
W = W_0 + Ae^{-aT} + Be^{-bT}
\end{equation}
where $a,b$ are constants and $A, B$ are constant in the original racetrack
model, but can be functions of fields in the case of D-term uplifting as in the
previous section. To include the open string modes then one adds the
additional term
\begin{equation}
W = W_{RT} + \lambda\phi\phi_+\phi_-
\end{equation}
Preliminary results suggest that cosmic strings are produced only during
inflation and are inflated away (Brax et al 2008). It is
therefore unlikely that cosmic strings are produced in the racetrack
models discussed here.

\section{Conclusions}
Recent developments in string theory have resulted in the
possibility of testing string theory with cosmology. Brane
inflation is the most promising string motivated model of
cosmological inflation. This predicts cosmic D-string formation at
the end of inflation, except for the possible case of moduli
inflation. It is thus possible that cosmic D-strings could provide
a window into physics at very high energy and very early times.
This exciting possibility has even been investigated in the
laboratory, see Haley, these proceedings, where the analogue of
branes in superfluid He-3 have been observed to annihilate and
leave remnant strings behind. Much work is still required before
realising the possibility of testing string theory with cosmology.
A better characterisation  of the properties of D-strings would be
necessary and a clear understanding of how their experimental
signals might differ from ordinary cosmic strings would be
required. Similarly, more theoretical work is necessary on string
motivated inflation models. However, the progress so far has been
fruitful.

\begin{acknowledgements}
This work is supported in part by STFC. ACD wishes to thank Tom Kibble and
George Pickett for the invitation to speak at this discussion meeting.
We wish to thank Stephen Davis, Rachel Jeannerot and Marieke Postma for
collaboration on some parts of this work. ACD thanks Fernando Quevedo for
fig 1.
\end{acknowledgements}

\label{lastpage}

\end{document}